%% file: power_law.tex
\documentclass{amsart}
\usepackage{amssymb, latexsym}
\pdfoutput=1
\usepackage{graphicx}
\usepackage{listings}
\begin{document}
\title{Random Processes with Power Law Spectral Density}
\author{Robert Kimberk}
\curraddr{R. Kimberk, Smithsonian Astrophysical Observatory}
\email{rkimberk@cfa.harvard.edu}
\author{Keara Carter}
\curraddr{K. Carter, Smithsonian Astrophysical Observatory}
\author{Todd Hunter}
\curraddr{T. Hunter, National Radio Astronomy Observatory}

\title{Random Processes with Power Law Spectral Density}

\begin{abstract}
A statistical model of discrete finite length random processes with negative power law spectral densities is presented. The definition of terms is followed by a description of the spectral density trend. An algorithmic construction of  random process, and a short block of computer code is given to implement the construction of the random process. The relationship between the second order properties of the random processes and the parameters of the construction is developed and demonstrated. The paper ends with a demonstration of the connection between the frequency of the random process sign changes and the power law exponent.
\end{abstract}

\maketitle

\section{Introduction}
This  paper provides a simple model of discrete random processes with negative  power law spectral densities, in what follows referred to as power law processes. This model is developed by deriving a construction, explaining the parameters, and demonstrating the predictive power the parameters have with respect to the second order properties of the random processes. The goal of the paper is to improve the understanding of power law processes such as martingales, fractionally integrated processes, and long memory processes.
This paper will include zero mean Gaussian processes with uniform spectral densities in the set of power law processes. It will be shown that  linear functions of a zero mean Gaussian random process can produce power law processes.  It will also be shown that the continuum of second order properties of power law processes start with Gaussian processes.

Experience, and literature demonstrate that power law processes are  manifest in a great many fields of study\cite{lb00} \cite{wp78} \cite{hh57}.   

\section{Definitions}

\subsection{Time Series}
This model will define a time series as a sequence of numbers that are discrete, real, and finite in both magnitude and sequence length. The $n$ elements of the sequence $x\sb{\tau}$ will be indexed by integers $\tau$, as in $0, 1,...,n-1$. The number of elements $n$ will be an even integer. The time between sequence elements will be one.

In order to improve the simplicity of exposition, except where noted, the sample mean will be calculated and subtracted from each element of the sequence to produce a zero mean sequence.
 The subtraction of the mean value will have no effect on the sample variance of the time series . This will allow the sample variance of the time series to be expressed as the mean value of the sum of the sequence elements squared as shown in the following expression.
 
 \begin{equation}
 Variance \   \  of \  \  Zero \  \ Mean\  \ Time \  \ Series. \   \ 1/n \sum_{\tau = 0}^{n-1} x\sb{\tau}^2
 \end{equation}

\subsection{Discrete Unitary Fourier Transform}
The unitary Fourier transform will be used in this model. The complex coefficients of the Fourier transform, $X \sb{\nu}$, will be indexed by values of  $\nu = -n/2 ... n/2$, incremented by 1, with the index $0$ dropped. The corresponding  Nyquist frequencies, cycles per sample, are  as  given by the sequence  $-1/2 , ..., -1/n, 1/n, ...,1/2$, as in \cite{jd53}. The zero frequency term, represented by $\nu = 0$, has been removed due to the mean subtraction. The plus and minus $1/2$ are the largest measurable  frequencies when sampled with a time period of one. The  frequency values plus and minus $1/n$ represent the smallest measurable frequencys of a zero mean sequence of length $n$. The forward and inverse unitary Fourier transforms are expressed by the following equations.

\begin{equation}
Forward \ \ Unitary \ \ Transform. \ \ X \sb {\nu} = n^{-1/2} \sum_{\tau = 0}^{n-1} x \sb{\tau} e^{-2 \pi i \tau \nu /n}
\end{equation}

\begin{equation}
Inverse\ \ Unitary \ \ Transform. \ \  x \sb{\tau} = n^{-1/2} \sum_{\nu = - n/2}^{n/2} X \sb{\nu} e^{2 \pi i \tau \nu /n}
\end{equation}

\subsection{Spectral Density} 
The coefficients  of the spectral density are the squared modulus of the complex unitary Fourier transform coefficients, $|X \sb{\nu}|^2$. The coefficients of the spectral density of a sequence of real numbers will be real, positive, and arrayed as an even function symmetrical about zero frequency. The relationship between spectral density and the variance of a zero mean process can be explained by using Plancherel's theorem applied to the unitary Fourier transform as expressed by the following equation.

\begin{equation}
Unitary \  \ Plancherel \  \ Theorem. \  \ \sum_{\tau = 0}^{n-1} x \sb{\tau}^2 = \sum_{\nu = - n/2}^{n/2} |X \sb{\nu}|^2
\end{equation}

Dividing both sides of the equation (2.4)  by $n$ yields the following relationship between the variance and spectral density.

\begin{equation}
Variance \ \ of \ Zero \  \ Mean \  \ Time \  \ Sequence. \  \  1/n \sum_{\tau = 0}^{n-1} x \sb{\tau}^2 = 1/n \sum_{\nu = - n/2}^{n/2} |X \sb{\nu}|^2
\end{equation}

The coefficients of the spectral density may also be generated using the Wiener-Khinchin theorem by taking the Fourier transform of the autocovariance function of a time series. Conversely the
autocovariance of a sequence is the inverse Fourier transform of the spectral density of the sequence.

\subsection{Convolution}
A discrete convolution is a type of multiplication between two sequences. There are two related versions, linear and circular. The linear discrete convolution of two sequences,f and g, is described in the following expression.

\begin{equation}
Discrete \  \ Linear \  \ Convolution. \  \  (f \ast g) \sb{i}  = \sum_{ k = -M}^{M} f \sb{k} g \sb{i-k}
\end{equation}

The circular convolution is illustrated by the following convolution of two sequences, $g = (1,2,3)$ and $ h = (4,5,6)$ to produce $ j = g \circledast h$. Sequence $g$ is first transformed into a circulant matrix.The left most column of the circulant matrix is the column vector of $g$, and the following columns are are circularly downshifted versions of $g$. The 3 by 3 matrix derived from $g$ is then multiplied with $h$ to produce $j$  on the right side of the equation that follows.

\begin{equation}
Circular \  \ Convolution. \  \  g \circledast h = j \   \   \
\begin{bmatrix}
1 & 3 & 2\\
2 & 1 & 3\\
3 & 2 & 1\\
\end{bmatrix}
\begin{bmatrix}
4\\
5\\
6\\
\end{bmatrix}
=
\begin{bmatrix}
31\\
31\\
28\\
\end{bmatrix}
\end{equation}

The result of the circular convolution may require a multiplicative factor, such as the $n^{-1/2}$ that precedes the unitary Fourier transform. This factor is required in the R language. To verify that  factor is correct  use the the Fourier convolution theorem: the Fourier transform of $g \circledast h$ should equal the product of the Fourier transforms of $g$ and $h$.

When one of the two factors in the convolution is a sequence of constants, the convolution may be called a moving average model. 

\section{The Power Law Trend of Spectral Density}
The spectral density of power law processes with real elements characteristically follow a power law trend, symmetric about zero frequency, as given by the following expression.

\begin{equation}
Spectral \  \ Density \  \ Trend. \  \   \  \ \alpha |f|^{- \beta}
\end{equation}

The term  $\alpha$ is a scaling factor greater than $0$, $|f|$ is the absolute value of the Fourier frequency sequence $-1/2 , ..., -1/n, 1/n,...,1/2$ , and $\beta$ is the power law exponent greater than or equal to zero. The width of the power law curve is controlled by $\beta$. When $\beta$ is equal to zero then the spectral density is uniformly flat, and as $\beta$ increases the width of the curve becomes more concentrated about zero. 

 Many of the interesting properties of power law processes follow from the spectral density trend. Estimates of the sample variance and sample variance of the mean of power law processes may be derived from the trend. The trend explains why, when $\beta$ is greater than zero, much of the effect of power law processes occurs at low frequencies, and equivalently, over long intervals of time. The Fourier similarity property combined with the Wiener-Khinchin theorem, found in section 2.3,  implies that as $\beta$ increases, and the spectral density becomes more narrow in frequency,  the autocovariance function widens. The widening of the autocovariance function increases the serial correlation of the power law sequence. The autocovariance derived from the power law spectral density trend will play a role in the construction of power law processes in the section that follows this section. The serial correlation will be measured by the frequency of sign changes of the zero mean power law process in the last section of this paper.

 Figure 1  is plot of expression (3.1) with $\alpha= 1$ and $\beta = 1$. 

\begin{figure} [h]
\begin{center}
\includegraphics [scale = .6] {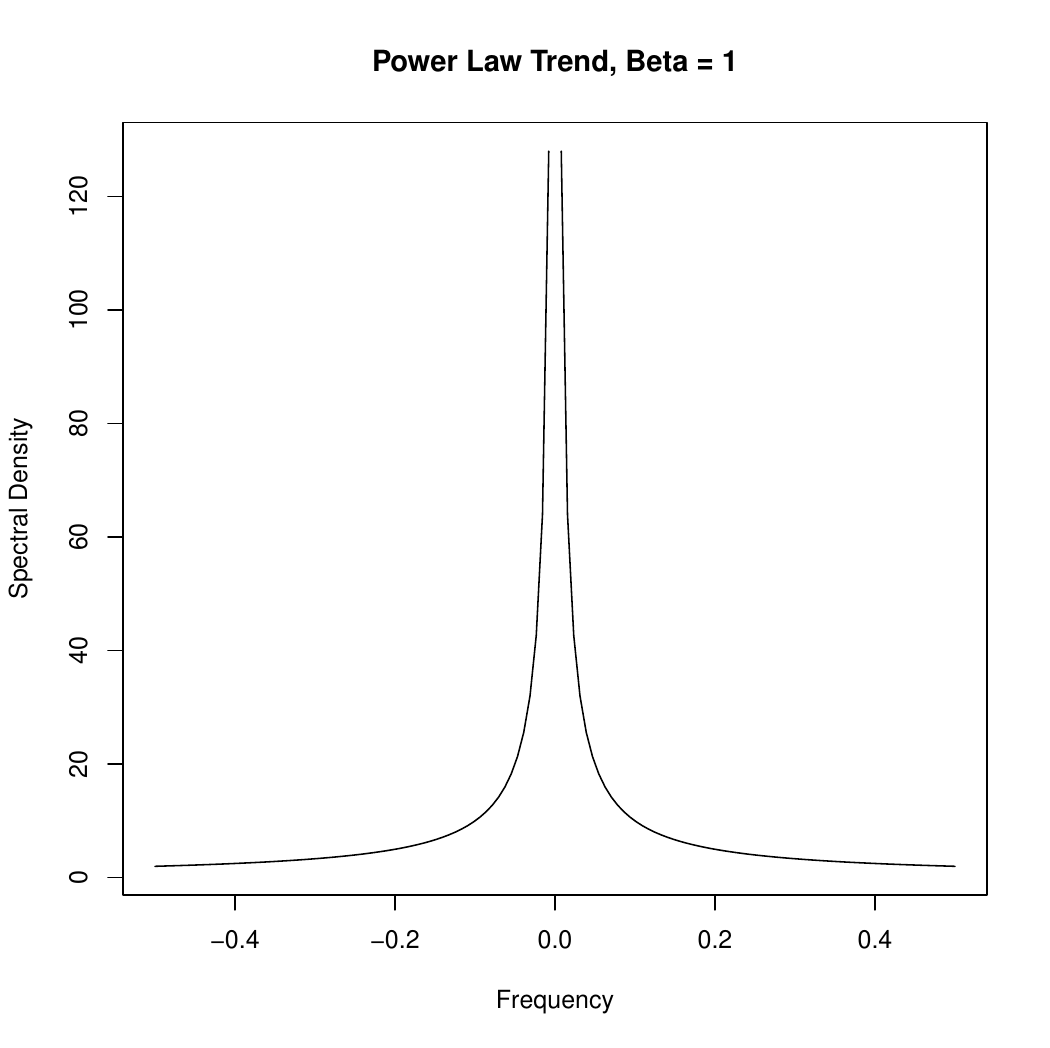}
\caption{Spectral Trend of a Power Law Process}
\label{ Fi : 1}
\end{center}
\end{figure} 

\section{A Construction of a Power Law Time Series Using Wold's Representation Theorem}
Herman Wold in 1938 \cite{hw38} \cite{ms83} developed a linear construction of any zero mean covariance stationary random processes. Removing an unneeded, for the purpose of this paper, deterministic term from Wold's decomposition yields the expression that follows.

\begin{equation}
Wold's \  \ Representation. \  \  \ x \sb{\tau} = \sum_{j=0}^{n-1} b \sb{j} \epsilon \sb{\tau - j}
\end{equation}

Equation (4.1) is the decomposition of the random sequence $ x \sb{\tau}$ into the linear convolution of a statistically uncorrelated, zero mean, random sequence $ \epsilon \sb{\tau}$ and a sequence of constants, $b \sb{j}$, that have a convergent sum. The linear or the circular convolution may be used in Wold's representation. 

\subsection{Circular Convolution Model for Power Law Sequences}   

This paper will use the circular convolution to construct random power law sequences. The following list provides the details the elements of the construction.

\begin{itemize}
\item The zero mean sequence, of length $n$, of Gaussian random numbers,  $\epsilon \sb{\tau}$,  with  variance equal to $\alpha$ in expression (3.1).
\item The elements of sequence $b \sb{j}$ are the modulus of the inverse Fourier transform coefficients of  the spectral density trend given in section 3, with $\beta$ divided by 2. The spectral trend is as follows:  $|f|^{-\beta/2}= (|-1/2|^{-\beta/2} ,..., \\ |-1/n|^{-\beta/2}, |1/n|^{-\beta/2}, ...,|1/2|^{-\beta/2})$. 
\item The power law time series $x \sb{\tau}$ of length n that is the product of the circular convolution $  b \sb{j} \circledast \epsilon \sb{\tau}$, and the factor $n^{-1/2}$.
\end{itemize}

\subsection{ The Model from Initial Sequences to Time Series to Spectral Density}

\begin{itemize}
\item Starting with the spectral trend with slope $-\beta/2$, the modulus of the inverse Fourier transform  is the autocovariance function as given by Wiener- Khinchin theorem, see section 2.3. This  autocovariance  sequence is $b \sb{j}$.
\item The power law time series is the circular convolution of, $b \sb{j}$, the autocovariance function of the spectral trend and the normal sequence $\epsilon \sb{\tau}$.
\item The Fourier Transform of the time series, due to the Fourier convolution theorem, is equal to the sequential product of the spectral trend with slope $-\beta / 2$  and the Fourier transform of the normal random sequence $\epsilon \sb{\tau}$.
\item Squaring the modulus of the Fourier transform of the power law time series produces the product of the spectral trend of slope $\beta$ and the spectral density of  $\epsilon \sb{\tau}$.
This product is the spectral density of the power law process with exponent $\beta$.
\end{itemize}

\subsection{How Wold's Model Works for Covariance Non-Stationary Power Law Sequences}  
Power law sequences, except when $\beta =0$, have a variance that increases as the sequence length increases, and are not covariance stationary. This paper's use of Wold's representation works for covariance non-stationary power law sequences because the sequence $b \sb{j}$ is recalculated as the length n of the sequence increases. The sequential recalculation of the power law sequence captures the effect of increasing variance as the length n increases. 

\subsection{R Language Code to Construct a  Power Law Time Sequence} 

 The following code will produce power law sequences $ x \sb{\tau}$  of arbitrary  $\alpha$ , $\beta$, and integer length $ n$.
 
\begin{lstlisting}
# Wold's construction of power law time series 
alpha = 25         	# variance of normal sequence
beta = 1.5         	# power law exponent				
n = 2e5 		# length of sequences				
set.seed(200)	# optional seed for normal sequence			
epsilon = rnorm(n, mean = 0, sd = alpha^0.5)  
frequency=c(seq(-1/2,-1/n,by=1/n),seq(1/n,1/2,by=1/n)) 
density=(abs(frequency))^(-beta/2)
b = Mod((n^(-1/2)) * fft( density, inverse = TRUE))
x = (n^(-1/2))*convolve(b,epsilon,type=``circular'')
\end{lstlisting}

\subsection{ Log-Log Graphs of Power Law Sequence Spectral Density}
Figure 2 is a pair of log-log graphs of  power law spectral densities. The time series for each plot was generated by the code in section 4.4. The plots demonstrate that the slope of the spectral density of a power law sequence has a linear trend when plotted in log-log format. The slope of the linear trend is equal to $-\beta$. In both plots the variance $\alpha$ of $\epsilon \sb{tau}$ is 25, and the length of each sequence is $2 \cdot 10^{3}$. The displayed frequency axis in both plots are the positive values of index $\nu$.

 \begin{figure}[h]
 \begin{center}
 \includegraphics [scale = .6] {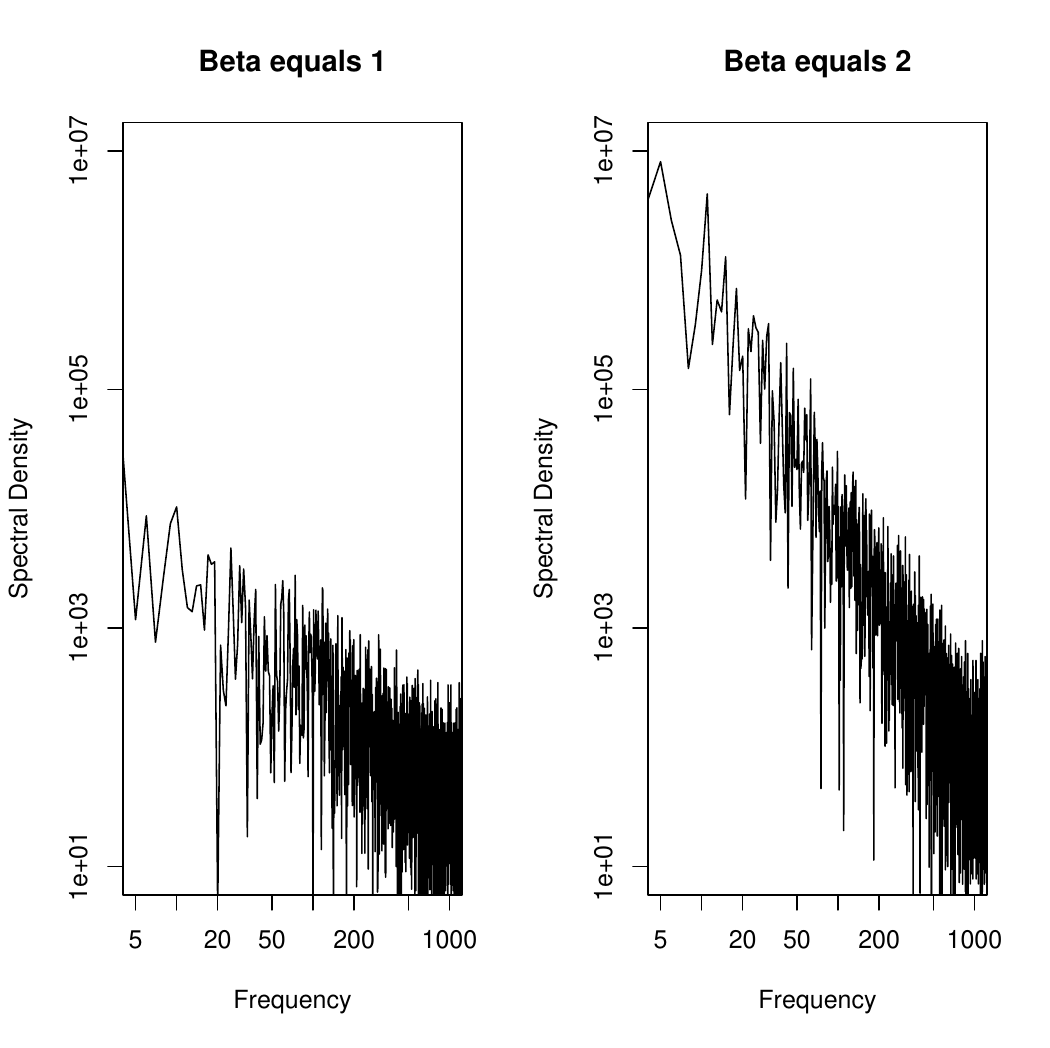}
 \caption{Spectral Density of Power Law Sequences Generated by Wold's Construction}
 \label{Fi : 2}
 \end{center}
 \end{figure}
 
\section{Estimated Sample Variance of a Power Law Process}
An estimate of the sample variance of a power law process may be derived from the parameters $\alpha, \beta$ and the length $n$ using the continuous integral of the expression in equation (3.1), \cite{dp85}. Equation (2.5) previously revealed that the sample variance of the power law time sequence is related to the spectral density. It is possible to think that the factor 1/n, in equation (2.5), is an approximation to a differential factor $df$, and that the right hand sum in equation (2.5) is an approximation of the continuous integral of spectral density. As the spectral density of a real process is an even function, symmetrical about zero, this paper will integrate the spectral density for positive frequencies, and multiply by two to calculate the integral over both positive and negative frequencies. 

Three examples of the technique will be given for value of $\beta$ equal to $ 0,1,2$, though the technique works for non-integer values of $\beta$. The three tables that follow will compare the estimate of sample variance derived from the integral of the spectral density function of parameters, and the sample variance calculated from a random process generated by the code in section 4.4. The computed power law sequences were generated with a constant seed for the pseudo-random Gaussian sequence $\epsilon \sb{\tau}$.

In what follows it should be remembered that $\alpha$ is the variance of $\epsilon \sb{\tau}$. In the following three tables the value of $\alpha$ will remain at 25.

\subsection{ White Process, $\beta = 0$}
This is the only power law process composed of uncorrelated random elements. It has a flat spectral density with an integral expressed as the following equation.

\begin{equation}
 White \  \ Process \  \ Variance. \  \  2 \alpha \int_{1/n} ^{1/2} f^{0} \  \  df = 2 \alpha \Bigr |_{1/n}^{1/2} f = 2 \alpha (1/2 -1/n) \approx \alpha
 \end{equation} 
 
 The white process has a constant sample variance independent from $n$. The following table displays the computed sample variance calculated from sequences produced by the code in section 4.4 versus the estimated sample variance using equation (5.1).  
 
 \begin{table} [h]
 \begin{center}
  \caption{Estimate of White Process Sample Variance} \label{Ta:1}
 \renewcommand{\arraystretch}{1.5}
 \begin{tabular}{  |  l  |  r  |  r  |  r  | }
 \hline
 $n$					& $\alpha$					& measured sample variance					& estimated sample variance  \\ \hline
 $2 \cdot 10^{3}$			& $25$					& $ 2.537 \cdot 10^{1} $						& $ \alpha =2.5 \cdot 10^{1} $ \\ \hline
 $2 \cdot 10^{4} $			& $25$					& $ 2.481 \cdot 10^{1} $						& $ \alpha =2.5 \cdot 10^{1} $ \\ \hline
 $ 2 \cdot 10^{5} $			& $25$					& $ 2.507 \cdot 10^{1} $						& $ \alpha =2.5 \cdot 10^{1} $ \\ \hline
 \end{tabular} 
 \end{center}
 \end{table}

\subsection{Flicker Process, $\beta = 1$}
The flicker process is composed of moderately serially correlated elements, and is logarithmically non-stationary with regard to variance. The integral of the spectral density is given as follows.

\begin{multline}
Flicker \  \ Process \  \ Variance. \  \  2 \alpha \int_{1/n}^{1/2} f^{-1} \  \ df  = 2 \alpha \Bigr|_{1/n}^{1/2} ln(f) \\
= 2 \alpha (ln(n) + ln(1/2)) \approx 2 \alpha  ln(n)
\end{multline}

The following table displays  the measured sample variance from the construction, and the estimate of sample variance from the integral of the spectral density.
\begin{table}[h]
\begin{center}
\caption{Estimate of Flicker Process Sample Variance} \label{Ta:2}
\renewcommand{\arraystretch}{1.5}
\begin{tabular}{  |  l  |  r  |  r  |  r  | }
\hline
 $n$					& $\alpha$					& measured sample variance					& estimated sample variance  \\ \hline
 $2 \cdot 10^{3}$			& $25$					& $3.585 \cdot 10^{2}$						& $2 \alpha ln (n) = 3.800 \cdot 10^{2}$ \\ \hline
 $2 \cdot 10^{4}$			& $25$					& $4.863 \cdot 10^{2}$						& $2 \alpha ln (n) = 4.952 \cdot 10^{2}$ \\ \hline
 $2 \cdot 10^{5}$			& $25$					& $6.030 \cdot 10^{2}$						& $2 \alpha ln (n) = 6.103 \cdot 10^{2}$ \\ \hline
 \end{tabular}

 \end{center}
 \end{table}
 
 \pagebreak

\subsection{Random Walk Process, $\beta = 2$}
The random walk process is composed of strongly serially  correlated elements, and is linearly non-stationary with respect to sample variance.The estimate of the sample variance from the integral of the spectral density is as follows.

\begin{multline}
Random \  \ Walk \  \ Process \  \ Variance \  \ 2 \alpha \int_{1/n}^{1/2} f^{-2} \  \  df \\ = 2 \alpha \Bigr|_{1/n}^{1/2} -f^{-1} = 2 \alpha (n-2) \approx 2 \alpha n
\end{multline}

The following table displays the computed sample variance from the sequence generated by the code in section 4.4 , and the estimated sample variance derived from the integral.

\begin{table}[h]
\begin{center}
\caption{Estimate of Random Walk Sample Variance} \label{Ta:3}
\renewcommand{\arraystretch}{1.5}
\begin{tabular}{ |  l  |  r  |  r  |  r  | }
\hline
$n$					& $\alpha$					& measured sample variance					& estimated sample variance  \\ \hline
$2 \cdot 10^{3}$			& $25$					& $ 1.313 \cdot 10^{5}$						& $ 2 \alpha n = 1 \cdot 10^{5} $ \\ \hline
$2 \cdot 10^{4}$			& $25$					& $ 1.374 \cdot 10^{6}$						& $ 2 \alpha n = 1 \cdot 10 ^{6} $ \\ \hline
$2 \cdot 10^{5}$			& $25$					& $ 1.028 \cdot  10^{7}$						& $ 2 \alpha n = 1 \cdot 10^{7} $ \\ \hline
\end{tabular}
\end{center}
\end{table}

\section{The Variance of the Sample Mean}
The variance of the sample mean of a power law sequence is another second order statistic that can be estimated from the parameters $\alpha, \beta$ and sequence length $n$. This section will start with a discussion of Markov's weak law of large numbers and its  relevance to the variance of the sample mean.

\subsection{Markov's Weak Law of Large Numbers}
Markov's weak law of large numbers provides a condition for the mean of an increasingly long sequence of a process to converge, in probability, to a fixed value. The law of large numbers is an important assumption in many statistical functions. However some finite discrete power law processes violate the law of large numbers, and Markov's law provides an explanation.

Markov's law, with no condition of statistical independence of the sequence elements \cite{bg62}, is given by the following expression where the notation  $Var(x \sb{n})$ is the variance of the sequence $x \sb{n}$.

\begin{equation}
Markov's \  \ Law. \  \  n^{-2} Var(\sum_{n=0}^ {n-1} x \sb{n})  \rightarrow 0, \  \ as \  \  n \rightarrow \infty
\end{equation} 

Markov's law may be algebraically transformed as a function of the variance of the sample mean as given by the next expression.

\begin{equation}
Markov's \  \ Law \  \  Transformed. \  \  Var(n^{-1} \sum_{n=0}^{n-1} x \sb{n}) \rightarrow 0, \  \  as \  \ n \rightarrow \infty
\end{equation}

The variance of the sample mean must asymptotically approach zero for the mean of a sequence to converge.

\subsection{Estimates of the Variance of the Sample Mean of Power Law Processes}
The zero frequency coefficient of the spectral density is associated with the mean value\cite{wf76}. This paper has avoided using the zero frequency coefficient of the spectral density. In the derivation of the estimate of the variance of the sample mean, the spectral density coefficient at frequency $1/n$ will be a surrogate for the zero frequency coefficient. The $1/n$ frequency term, evaluated in expression (3.1) will form the estimate of the variance of the mean. The product of the surrogate differential $1/n$, and the spectral density at $1/n$ frequency, forms the estimate as shown in the following development of the expression $ \alpha n^{\beta -1}$.

\begin{multline}
Variance \  \ of  \  \ the \  \ Sample \  \ Mean. \  \  1/n \  \ \alpha |f|^{- \beta} = 1/n\  \ \alpha |1/n|^{- \beta}= \alpha n^{\beta -1}
\end{multline}

In the table that follows the constructed power law sequences do not have the mean value subtracted from the sequence elements. Using the construction code of section 4.4 to generate 100 power law  sequences, of length $ n = 2 \cdot 10^5$, and $\alpha = 25$, for each value of $\beta$, a sample variance of the mean for each $\beta$ was calculated. The calculated variance of the mean is compared to the estimate produced by the function $ \alpha n^{\beta -1}$.

\begin{table}[h]
\begin{center}
\caption{Estimate of Sample Variance of the Mean}  \label{Ta:4}
\renewcommand{\arraystretch}{1.5}
\begin{tabular} {  |  l  |  r  |  r  |  r  | }
\hline
$\beta$			& calculated variance of the mean		& estimated variance of the mean		& estimate function \\  \hline
$0$				& $1.431 \cdot 10^{-4}$					& $ 1.25 \cdot 10^{-4}$				& $ \alpha n^{-1}$ \\ \hline
$1$				& $ 2.479 \cdot 10^{1}$					& $ 2.5 \cdot 10^{1}$				& $ \alpha n^{0} $ \\ \hline
$2$				& $ 5.797 \cdot 10^{6}$					& $  5.0 \cdot 10^{6}$				& $ \alpha n $ \\ \hline
\end{tabular}
\end{center}
\end{table}

The results of the table suggest that the variance of the sample mean of a power law sequence, with $\beta= 0$ will decrease with increasing sequence length $n$.
The variance of the sample mean of a power law sequence with $\beta =1$ will have a constant variance of $\alpha$. The variance of the sample mean of a power law sequence
with $\beta= 2$ will have a nonstationary mean with increasingly large variance.  

\section{The Probability of a Sign Change in a Zero Mean Power Law Sequence}
  R. C. Geary's 1970 paper that describes using the frequency of sign changes to detect serial correlation is the inspiration for this section, \cite{rg70}. Starting with a zero mean power law sequence, at each increment of the index $\tau$ of the sequence $x_{\tau}$, a measure of a sign change is a Bernoulli trial. The expected number of sign changes in a sequence of length $n$, is $n-1$ times the probability of a sign change $p$, with $np$ being the sample mean of a binomial distribution. When $\beta = 0$, and the sequence is serially uncorrelated, then $ p \approx 0.5$, and as $\beta$ increases the value of $p$ decreases due to the increasing serial correlation of the power law sequence.
  
  Table 5 is the result of a numerical test of whether the probability of sign changes would be a simple method for measuring the $\beta$ of a zero mean power law sequence. Sequences of length 1024 were generated using the code in section 4.4, and the number of measured sign changes was divided by 1023 to estimate $p$, the probability of sign changes.

\begin{table}[h]
\begin{center}
\caption{Probability of Sign Change as Measure of Beta} \label{Ta:5}
\renewcommand{\arraystretch}{1.5}
\begin{tabular} { |  l  |  r  |  r  | }
\hline
$\beta$			& number of sign changes			& probability of sign change \\ \hline
$0$				& $494$					& $0.48$ \\ \hline
$0.2$				& $460$					& $0.45$ \\ \hline
$0.4$				& $436$					& $0.43$ \\ \hline
$0.6$				& $380$					& $0.37$ \\ \hline
$0.8$				& $332$					& $0.32$ \\ \hline
$1.0$				& $258$					& $0.25$ \\ \hline
$1.2$				& $202$					& $0.20$ \\ \hline
$1.4$				& $124$					& $0.12$ \\ \hline
$1.6$				& $90$					& $0.09$ \\ \hline
$1.8$				& $68$					& $0.07$ \\ \hline
$2.0$				& $26$					& $0.03$ \\ \hline
\end{tabular}
\end{center}
\end{table}

\input{power_law.bbl}

\end{document}

%% file: power_law.bbl
\pdfoutput=1